\documentstyle[12pt,epsf]{article}
\newcommand{\be}{\begin{equation}}
\newcommand{\ee}{\end{equation}}
\newcommand{\bea}{\begin{eqnarray}}
\newcommand{\eea}{\end{eqnarray}}
\newcommand{\ci}{\cite}
\newcommand{\bi}{\bibitem}
\newcommand{\nono}{\nonumber \\}

\newcommand{\dd}{\partial}

\newcommand{\half}{\frac{1}{2}}

\def\dal{\,\lower0.3ex\vbox{\hrule\hbox{\vrule\kern2pt\vbox{\kern4pt\kern4pt}
\kern2pt\vrule}\hrule}\,}

\def\L{{\cal L}}

\begin{document}
\title{{\sl Trapping of a model-system for a soliton in a well}}
\vspace{1 true cm}
\author{G. K\"albermann\\
Soil and Water department\\
Faculty of Agriculture\\ Rehovot 76100
, Israel}

\maketitle

\begin{abstract}
\baselineskip 1.5 pc
The nature of the interaction of a soliton with an attractive 
well is elucidated using
a model of two interacting point particles.
The system shows the existence of trapped states
at positive kinetic energy, as well as reflection by
an attractive impurity, as found when a topological soliton
scatters off an attractive well. 
\end{abstract}

%\newpage
%\tableofcontents
\newpage
\baselineskip 1.5 pc

\section{\label{introduction}\sl Introduction}

Topological solitons arise as nontrivial solutions
in field theories with nonlinear interactions.
These solutions are stable against dispersion. 
Topology enters through the absolute conservation of a topological
charge, or winding number.\ci{raj}

It is for this reason they become so important in the description
of phenomena like, optical self-focusing,
magnetic flux in Josephson junctions\ci{shen}
or even the very existence of stable elementary
particles, such as the skyrmion \ci{sk,an}, as a model of hadrons.

Interactions of solitons with external agents become
extremely important. These interactions
allow us to test the validity of such models in real situations.

In a previous work \ci{kal97} the interaction of a 
soliton in one space dimension with 
finite size impurities was investigated. 

In the works of Kivshar et al.\ci{kivshar} (see also ref.~\ci{cuba,cuba1}),
it was found that the soliton displays
unique phenomena when it interacts with an external impurity.
The existence of trapped solutions
for positive energy or, bound states in the continuum, is
a very distinctive effect for the soliton in interaction
with an attractive well.

We can understand the origin of impurity interactions
of a soliton by looking at the impurity as a nontrivial
medium in which the soliton propagates.
An easy way to visualize these interactions consists in 
introducing a nontrivial metric
for the relevant spacetime. 
The metric carries the
information of the medium characteristics.

A 1+1 dimensional scalar field theory 
supporting topological solitons in flat space, 
immersed in a backgound determined by the metric $g_{\mu\nu}$ in
a minimal coupling to the metric, is given by

\be\label{lag}
\L  = \sqrt{g}\bigg[g^{\mu\nu}\half \dd_\mu\phi\,\dd_\nu\phi
- U(\phi)\big]
\ee
where $g$ is the of the determinant of the metric, and
U is the self-interaction that enables the existence of
the soliton. For a weak potential we have\ci{rob}

\bea\label{metric}
{g_{00}}&\approx&1+V(x)\nono
{g_{11}}&=&-1\nono
{g_{-11}}&=&g_{1-1}=0
\eea

Where $V(x)$ is the external space dependent potential.
The equation of motion of the soliton in the background space becomes

\be\label{eq1}
\frac{{\dd}^2 \phi}{\dd t^2} -{\sqrt{g}}^{-1}
 \frac{\dd}{\dd x}\big[\sqrt{g}\frac{\dd\phi}{\dd x}\bigg]\nono 
+ g_{00}\frac{\dd U}{\dd\phi}=0.
\ee

This equation is identical, for slowly varying
potentials, to the equation of motion of a soliton interacting with
an impurity $V(x)$. 
Impurity interactions are therefore acceptable couplings of
a soliton to an external potential.
It is also the only way to couple the soliton without spoiling
the topological boundary conditions.
The source term generated by the metric is essentially
a space dependent mass term.

The interaction of a soliton with an attractive impurity
shows, however, some puzzling effects\ci{kivshar,kal97}.
A soliton can be trapped in it, when it impinges onto the well
with positive kinetic energy. 
Energy conservation demands that
the soliton fluctuates and distorts in trapped states inside the well.
Even more counterintuitive is the fact that the soliton can be reflected
by the well.

Neither of these effects are possible for classical
point particles.
The difference must obviously be due to the extended
character of the soliton. 

This was indeed put in evidence in the works of Kivshar et al.\ci{kivshar}.
It was shown there that the bulk soliton behavior may be reproduced
qualitatively, by
taking the center of the soliton as a time dependent collective
coordinate coupled to the major excitation modes of the soliton in the
well, the impurity mode, and a shape distortion mode.
The first mode is excited inside the well only and appears as
an oscillating packet centered at the well. The shape mode is an
excitation that accompanies
the soliton with a time dependent amplitude along its scattering,
and accounts for the distortions of
the free soliton. 
The dynamics of the soliton was then replaced by the
equations of motion of three classical particle-like excitations: the
center of the soliton, the amplitude of the impurity mode and
the amplitude of the shape mode. 
For a $\delta$ function type of well, it was found that the system behaves
analogously to the soliton. Kinetic energy of the
soliton center can be transferred to the impurity and the shape modes. The
system of three effective degrees of freedom can resonate inside
the well and be trapped. Reflection by the attractive
well is also observed. 
The system captures the essential features of the behavior of the soliton.

However, both the interactions between the collective
degrees of freedom and the dynamics of the system are derived
from the soliton itself. Moreover, the behavior
was shown to hold for $\delta$ type of wells only.

Later it was shown \ci{kal97} that all the
features of the scattering of a soliton show up
in the case of a fine size well too.

In the present work we show that the effects are not specific to the
soliton generated dynamics and interactions.

We are able to reproduce here all the abovementioned effects with a 
classical model for an extended object, regardless of the
soliton dynamics.
In the next section we show that a simple classical model exhibits the 
same behavior as the soliton. It can be trapped and reflected
by an attractive well.
The system will also show chaotic behavior.

The classical system can serve as a nice introductory
example for the surprising behavior of solitons. It demands 
only basic lagrangian mechanics knowledge, but, it has
many interesting features that are easily visualized
by undergraduate students. It may serve also as an example
of chaotic behavior in classical mechanics.

\section{\label{trap}\sl A classical model of trapping}

Kinks and other topological deffects are sometimes used
to model classical systems of masses. We will here find, that
the analogy is more than mathematical. The very behavior of the
soliton is exactly the one observed on a two-body
system interacting with an external potential.

In ref.\ci{kivshar}, the dynamics of a soliton interacting with
a well was also studied by selecting the collective coordinates
of the soliton center, its shape-mode excitation amplitude and
an impurity mode. Using this scheme, it was found that this
classical system shows an analogous behavior
to the soliton itself. Namely, trapped states and reflection by the
attractive impurity. 
The specific dynamics of the soliton in terms of its collective coordinate,
and its excitations including the shape mode were crucial for 
the obtention of the abovementioned results.

We show here that the effects are generic, any system resembling
the behavior of the soliton, mainly, its extended character, will
indeed yield similar results.
This is an unexpected situation whose motivation arose entirely
from the behavior of the soliton and its relevance goes beyond it,
as will be explained below.
 
In order to justify the connection with
the behavior of the model we appeal to a physical scenario
in which both Sine-Gordon solitons and Kinks arise.\ci{rem}
Sine-Gordon solitons arise in large Josephson junctions
and in the motion of dislocations in a one-dimensional crystal.
Kinks arise in the latter also when the substrate potential is
nonperiodic.

Consider the Hamiltonian of dislocations in a crystal with
nearest neighbor interactions \ci{rem}

\bea\label{disloc}
H=\sum\bigg[\half~m~\big(\frac{du_n}{dt}\big)^2+\frac{G}{2}~
\big(u_{n+1}-u_n\big)^2+~V(u_n)\bigg]
\eea

Where $u_n$ is the displacement of the dislocation at site $\sl{n}$, 
$G$ is the spring constant between particles and $V$ is a
site potential generated by the substrate chain of fixed
particles upon which the mobile dislocations move.

The above Hamiltonian supports solitons in the continuum, strong
coupling limit. In particular for kinks, only a few 
dislocation centers are needed to generate
the desired effect of the moving soliton.
The minimal set would then be a couple of dislocation 'particles'
moving along the substrate.
Now suppose the above model is applied to a substrate for
which the parameters of the substrate potential vary.
This is analogous to the variation of the metric
in the description of the previous section.
In such a scenario we have to modify the substrate potential
by adding a local interaction at fixed sites in the
lattice.
This is essentially the procedure in soliton-impurity 
scattering. If the effects found by Kivshar et al.\ci{kivshar}, 
are indeed based on the above simple dislocation model, then these should
appear clearly when a couple of dislocation degrees of freedom
scatter off an external potential.
We will see below that this borne out in a simple two-particle
model that imitates the soliton behavior.

Consider a system of two classical point particles connected by
a massless spring, two of the dislocations above, and
a repulsive force between them needed to prevent
their collapse to zero size that subsums the behavior of the rest of the chain
of dislocations. With only two degrees of freedom, 
we are eliminating the rigidity of the chain, thereby introducing
the spurious possibility of complete overlap between the two sites,
which does not occur when the chain is infinite. 
Hence the need for a repulsive interaction.

The above simplistic model 
is not directly related to the collective coordinate treatment
of ref.\ci{kivshar} deliberately. 
The aim is to show that the particular
dynamics of the soliton is {\sl not} the cause of the peculiar phenomena
previously found.

When each particle in the system is allowed to interact 
with an external potential, we are imitating the impurity
force or the local change in the dislocation potential.

The classical nonrelativistic one-dimensional lagrangian for
the system of equal masses $m_1=m_2=1$ becomes:
\be\label{clas1}
{\L}_{sys}  = \frac{\dot{x}_1^2}{2}+\frac{\dot{x}_2^2}{2}-
k~\frac{(x_1-x_2)^2}{2}-\frac{\alpha}{{|x_1-x_2|}^n}
+V(x_1)+V(x_2)
\ee
For the potential well we take
\be\label{well}
V(x)~=A~e^{-\beta~x^2}
\ee
Although any finite size well may serve for this purpose.
We prepare the two-particle system at rest at a large distance
far away from the well with  an initial speed $v$. 
The equilibrium interparticle separation is ${r_0}^{n+2}~=~\frac
{n~\alpha}{k}$. We here use $n=2$.

The equations of motion are not solvable analytically.
However we can show that for a well large compared to
the equilibrium distance $r_0$ the system may be trapped and
oscillate inside it.
Transforming to relative and center of mass coordinates, 
$r=\frac{x_2-x_1}{2}~,~R=\frac{x_1+x_2}{2}$ and using
the ansatz $r=r_0+\delta(t)$, with $\delta$ a small parameter, we
find the equations of motion near the center of the well $R=0$

\bea\label{approx}
\ddot{\delta}~+2~k~\delta+2~A~e^{-r_0^2~\beta}~\beta~(r_0+\delta)=0\nono
\ddot{R}+2~A~\beta~R~e^{-r_0^2~\beta}=0
\eea

Where we have used $\beta~r_0^2<< 1$, a wide well as compared
to the equilibrium distance of the system.
Passing to a new coordinate

\bea\label{transf}
\delta(t)=-\frac{r_0}{1+\frac{k}{A~\beta}~e^{\beta~r_0^2}}+~\epsilon(t)
\eea

the first of equations \ref{approx} becomes

\be\label{approx1}
\ddot{\epsilon}~+2~k~\epsilon+2~A~e^{-r_0^2~\beta}~\beta\epsilon=0
\ee

It is clear from the above equations that the center of mass coordinate
of the system oscillates around the center of the well, while
the relative coordinate oscillates too with the small amplitude
$\epsilon$. Moreover the system shrinks inside the well.
The oscillations of the center of mass coordinate $r$ compensate 
for the loss of kinetic energy of the
system that impinged from infinity with a fixed relative separation.
The above treatment demonstrates, that at least there is room
for the trapping to occur.

We now proceed to show the numerical
results of the exact calculation of the development of the model.

Using the numerical method used in ref.~{\ci{kal97}}
we can find the outcome of the scattering events as a 
function of the initial speed.
The system starts with a certain initial center of mass velocity 
$v$ far away from the well. The system is prepared with 
the relative separation of equilibrium $r_0$ and the outcome
of the scattering is monitored as a function of the initial 
conditions.

Figure 1 exemplifies the results
for the choice of parameters $k=1,~\alpha=1,~n=2,~A=2,~\beta=1$.

Quite unexpectedly, it is found that the system behaves exactly
like the soliton.

The system can be trapped $v_{final}=~0$, reflected, $v_{final}<~0$ 
or transmitted, $v_{final}>~0$ through the well
 by varying the initial speed. 

When the system is trapped, it oscillates with a null 
average speed, the kinetic energy stored in the vibrational and
deformation modes.

Minute changes of the initial speed around a value leading to a trapped state,
may generate reflection or transmission events.

The effects are independent of
the functional dependence of the interactions and external potential,
as well as the values of the parameters. 

\begin{figure}[tb]
\epsffile{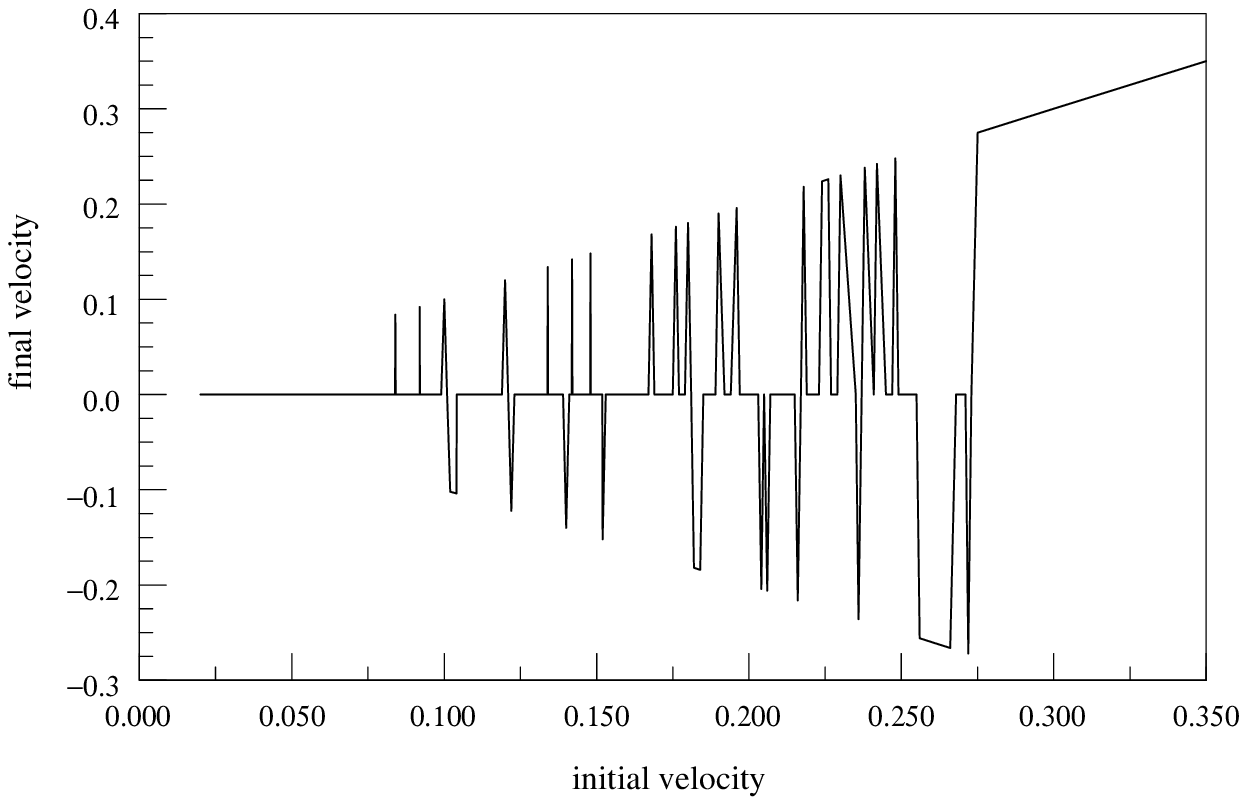}
\vsize=5 cm
\caption{\sl{Final velocity of the two-particle system 
as a function of the initial velocity
for the parameters $k=1,~\alpha=1,~n=2,~A=2,~\beta=1$ with
a velocity grid dv=.001}}
\label{fig1}
\end{figure}

In figure 1 we used a grid for $v$ of $dv=.001$. 
Using a finer
grid, each region of reflection-transmission unfolds to more islands
of trapping, reflection and transmission.

Finer and finer grids show more and more structure. 

Figure 2 shows a detailed expansion of the velocity range around v=.12 with 
a grid dv=.0002. 
The system is chaotic, an infinitesimal change in the initial speed
produces diverging results.

Many of the phenomena related to chaotic behavior may
be identified in the system, namely scaling, bifurcation 
and perhaps even fractal structure.

\begin{figure}[tb]
\epsffile{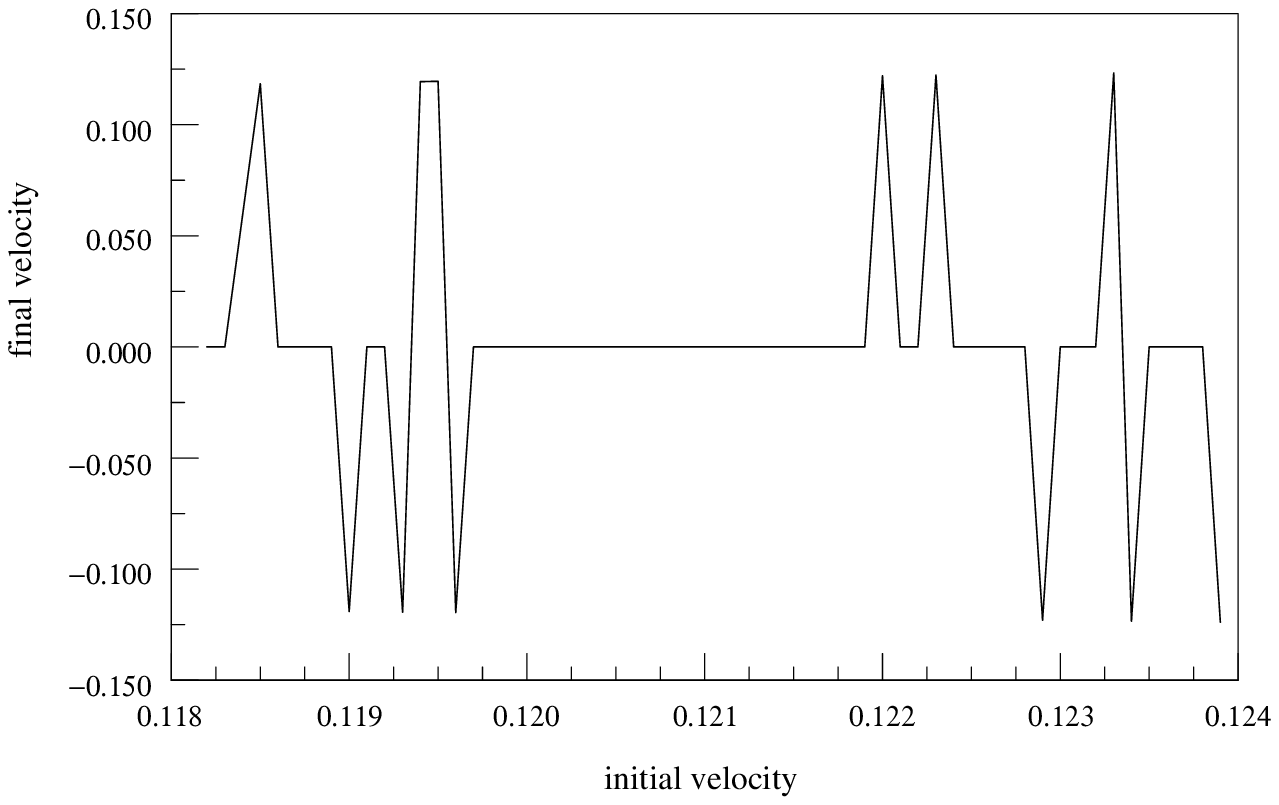}
\caption{\sl{Same as figure 1, but with finer velocity grid
dv = .0002}}
\label{fig2}
\end{figure}

It is now safer to claim that the unexpected behavior of a soliton interacting 
with an attractive well may be traced back to its extended nature.
If we regard each ${\phi}(x)$ as a classical pointlike
object we will find interactions between neighboring particles
of attractive and repulsive character. The basic attractive interaction is
provided by the space derivative of the soliton lagrangian and 
a piece of the self-interaction potential, while the repulsive 
interaction is provided by the latter and the coupling to the
remainder of the soliton or linear chain ina discrete model.

\section{\sl Final remarks}

The simplest implementation of the system studied
would be a toy-like system of
two masses tied-up to a spring sliding on a frictionless table
with a carved well on it. 
Two atoms in a molecule 
scattering off an external Van-der-Waals potential might
show the same effects in a quasi-classical approximation. 
However, quantum effects can blurr the
picture due to interference.

Turning the process backwards: an extended object,
may it be a soliton or a classical assembly of bound particles, in a trapped
state, can suddenly be freed from it provided some random
interaction causes the reversal of the process of trapping,
a process reminiscent of the decay of metastable states in quantum
mechanics. 
The concept of trapping in general is not discussed in the
classical mechanics literature. 
Soliton physics has taught us that an extended object made
of particles in interaction has a much richer behavior than
expected.

{\bf Acknowledgements}

This work was supported in part by the Department of
Energy under grant DE-FG03-93ER40773 and by the National Science Foundation
under grant PHY-9413872.

\end{document}